\documentclass[12pt] {article}
\begin{document}
\setcounter{page}{1}
\def\theequation{\arabic{section}.\arabic{equation}}
\def\theequation{\thesection.\arabic{equation}}
\setcounter{section}{0}

\title{On the $\Delta\Delta$ component of the deuteron in
the relativistic field theory model of the deuteron}

\author{A. N. Ivanov~\thanks{E--mail: ivanov@kph.tuwien.ac.at, Tel.:
+43--1--58801--14261, Fax: +43--1--58801--14299}~${\textstyle
^\ddagger}$, H. Oberhummer~\thanks{E--mail: ohu@kph.tuwien.ac.at,
Tel.: +43--1--58801--14251, Fax: +43--1--58801--14299} ,
N. I. Troitskaya~\thanks{Permanent Address: State Technical
University, Department of Nuclear Physics, 195251 St. Petersburg,
Russian Federation} , M. Faber~\thanks{E--mail:
faber@kph.tuwien.ac.at, Tel.: +43--1--58801--14261, Fax:
+43--1--58801--14299}}

\date{\today}

\maketitle

\begin{center}
{\it Institut f\"ur Kernphysik, Technische Universit\"at Wien,\\
Wiedner Hauptstr. 8-10, A-1040 Vienna, Austria}
\end{center}

\begin{center}
\begin{abstract}
The $\Delta\Delta$ component of the deuteron, where $\Delta$ stands
for the $\Delta(1232)$ resonance, is calculated in the relativistic
field theory model of the deuteron. For the probability of the
$\Delta\Delta$ component of the deuteron we give $P(\Delta\Delta) =
0.08\,\%$. This prediction agrees good with the experimental estimate
$P(\Delta\Delta) < 0.4\,\%$ at 90$\%$ of CL (D. Allasia {\it et al.},
Phys. Lett. B174 (1986) 450).
\end{abstract}
\end{center}

\begin{center}

\end{center}

\newpage

\section{Introduction}
\setcounter{equation}{0}

\hspace{0.2in} As has been stated in Ref.[1] that nowadays there is a
consensus concerning the existence of non--nucleonic degrees of
freedom in nuclei. The non--nucleonic degrees of freedom can be
described either within QCD in terms of quarks and gluons [2] or in
terms of mesons and nucleon resonances [3].

In this letter we investigate the non--nucleonic degrees of freedom in
terms of the $\Delta(1232)$ resonance and calculate the contribution
of the $\Delta\Delta$ component of the deuteron in the
relativistic field theory model of the deuteron (RFMD) [4--10].  As
has been shown in Ref.[7] the RFMD is motivated by QCD.  The deuteron
appears as a neutron--proton collective excitation, the Cooper
np--pair, induced by a phenomenological local four--nucleon
interaction in the nuclear phase of QCD. The RFMD describes the
deuteron coupled to hadrons through one--nucleon loop exchanges
providing a minimal transfer of nucleon flavours from initial to final
nuclear states and accounting for contributions of nucleon--loop
anomalies which are completely determined by one--nucleon loop
diagrams. The dominance of contributions of nucleon--loop anomalies to
effective Lagrangians of low--energy nuclear interactions is justified
in the large $N_C$ expansion, where $N_C$ is the number of quark
colours [7]. As has been shown in Refs.[8,9] the RFMD describes very good
the processes of astrophysical interest such as the neutron--proton
radiative capture n + p $\to$ D + $\gamma$, disintegration of the
deuteron by anti--neutrinos and neutrinos caused by $\bar{\nu}_{\rm
e}$ + D $\to$ e$^+$ + n + n charged and $\bar{\nu}_{\rm e}$ + D $\to$
$\bar{\nu}_{\rm e}$ + n + p neutral weak currents [8], the solar
proton burning p + p $\to$ D + e$^+$ + $\nu_{\rm e}$ and the
pep--process p + e$^-$ + p $\to$ D + $\nu_{\rm e}$ [9].

The phenomenological ${\rm n p D}$ interaction is given by [4--9]
the phenomenological Lagrangian
\begin{eqnarray}\label{label1.1}
{\cal L}_{\rm npD}(x) = - ig_{\rm V}[\bar{p}(x)\gamma^{\mu}n^c(x) -
\bar{n}(x)\gamma^{\mu}p^c(x)] D_{\mu}(x),
\end{eqnarray}
where $D_{\mu}(x)$, $n(x)$ and $p(x)$ are the interpolating fields of
the deuteron, the neutron and the proton [4--9]. The phenomenological
coupling constant $g_{\rm V}$ is related to the electric quadrupole
moment of the deuteron $Q_{\rm D} = 0.286\,{\rm fm}$: $g^2_{\rm V} =
2\pi^2 Q_{\rm D}M^2_{\rm N}$ [4,7], where $M_{\rm N} = 940\,{\rm MeV}$
is the nucleon mass. In the isotopically invariant form the
phenomenological interaction  Eq.(\ref{label1.1}) can be written as
\begin{eqnarray}\label{label1.2}
{\cal L}_{\rm npD}(x) = g_{\rm
V}\,\bar{N}(x)\gamma^{\mu}\tau_2N^c(x)\,D_{\mu}(x),
\end{eqnarray}
where $\tau_2$ is the Pauli isotopical matrix and $N(x)$ is a doublet
of a nucleon field with components $N(x) = (p(x), n(x))$, $N^c(x) =
C\,\bar{N}^T(x)$ and $\bar{N^c}(x) = N^T(x)\,C$, where $C$ is a charge
conjugation matrix and $T$ is a transposition.

In the RFMD [5,8] the $\Delta(1232)$ resonance is the
Rarita--Schwinger field [10] $\Delta^a_{\mu}(x)$, the isotopical index
$a$ runs over $a = 1,2,3$, having the following free Lagrangian
[11,12]:
\begin{eqnarray}\label{label1.3}
\hspace{-0.5in} {\cal L}^{\Delta}_{\rm kin}(x) =
\bar{\Delta}^a_{\mu}(x) [-(i\gamma^{\alpha}\partial_{\alpha} -
M_{\Delta}) \,g^{\mu\nu} +
\frac{1}{4}\gamma^{\mu}\gamma^{\beta}(i\gamma^{\alpha}\partial_{\alpha}
- M_{\Delta}) \gamma_{\beta}\gamma^{\nu}] \Delta^a_{\nu}(x),
\end{eqnarray}
where $M_{\Delta} = 1232\,{\rm MeV}$ is the mass of the $\Delta(1232)$
resonance field $\Delta^a_{\mu}(x)$. In terms of the eigenstates of
the electric charge operator the fields $\Delta^a_{\mu}(x)$ are given
by [11,12]
\begin{eqnarray}\label{label1.4}
\begin{array}{llcl}
&&\Delta^1_{\mu}(x) = \frac{1}{\sqrt{2}}\Biggr(\begin{array}{c}
\Delta^{++}_{\mu}(x) - \Delta^0_{\mu}(x)/\sqrt{3} \\
\Delta^+_{\mu}(x)/\sqrt{3} - \Delta^-_{\mu}(x)
\end{array}\Biggl)\,,\,
\Delta^2_{\mu}(x) = \frac{i}{\sqrt{2}}\Biggr(\begin{array}{c}
\Delta^{++}_{\mu}(x) + \Delta^0_{\mu}(x)/\sqrt{3} \\
\Delta^+_{\mu}(x)/\sqrt{3} + \Delta^-_{\mu}(x)
\end{array}\Biggl)\,,\\
&&\Delta^3_{\mu}(x) = -\sqrt{\frac{2}{3}}\Biggr(\begin{array}{c}
\Delta^+_{\mu}(x) \\ \Delta^0_{\mu}(x) \end{array}\Biggl).
\end{array}
\end{eqnarray}
The fields $\Delta^a_{\mu}(x)$ obey the subsidiary constraints:
$\partial^{\mu}\Delta^a_{\mu}(x) = \gamma^{\mu}\Delta^a_{\mu}(x) = 0$
[10--12]. The Green function of the free $\Delta$--field is determined by
\begin{eqnarray}\label{label1.5}
\hspace{-0.5in}<0|{\rm T}(\Delta_{\mu}(x_1)\bar{\Delta}_{\nu}(x_2))|0>
= - i S_{\mu\nu}(x_1 - x_2).
\end{eqnarray}
In the momentum representation $S_{\mu\nu}(x)$ reads [5,8,11,12]:
\begin{eqnarray}\label{label1.6}
\hspace{-0.5in}S_{\mu\nu}(p) = \frac{1}{M_{\Delta} - \hat{p}}\Bigg( -
g_{\mu\nu} + \frac{1}{3}\gamma_{\mu}\gamma_{\nu} +
\frac{1}{3}\frac{\gamma_{\mu}p_{\nu} -
\gamma_{\nu}p_{\mu}}{M_{\Delta}} +
\frac{2}{3}\frac{p_{\mu}p_{\nu}}{M^2_{\Delta}}\Bigg).
\end{eqnarray}
The most general form of the ${\rm \pi N \Delta}$ interaction
compatible with the requirements of chiral symmetry reads [11]:
\begin{eqnarray}\label{label1.7}
\hspace{-0.5in}&&{\cal L}_{\rm \pi N \Delta}(x) = \frac{g_{\rm \pi
N\Delta}}{2M_{\rm
N}}\bar{\Delta}^a_{\omega}(x)\Theta^{\omega\varphi}N(x)
\partial_{\varphi}\pi^a(x)
+ {\rm h.c.} = \nonumber\\
\hspace{-0.5in}&&= \frac{g_{\rm \pi N\Delta}}{\sqrt{6}M_{\rm
N}}\Bigg[\frac{1}{\sqrt{2}}\bar{\Delta}^+_{\omega}(x)\Theta^{\omega\varphi}
n(x) \partial_{\varphi}\pi^+(x) -
\frac{1}{\sqrt{2}}\bar{\Delta}^0_{\omega}(x)\Theta^{\omega\varphi}
p(x) \partial_{\varphi}\pi^-(x)\nonumber\\
\hspace{-0.5in}&&- \bar{\Delta}^+_{\omega}(x)\Theta^{\omega\varphi}
p(x) \partial_{\varphi}\pi^0(x) -
\bar{\Delta}^0_{\omega}(x)\Theta^{\omega\varphi} p(x)
\partial_{\varphi}\pi^0(x) + \ldots \Bigg],
\end{eqnarray}
where $\pi^a(x)$ is the pion field with the components $\pi^1(x) =
(\pi^-(x) + \pi^+(x))/\sqrt{2}$, $\pi^2(x) = (\pi^-(x) -
\pi^+(x))/i\sqrt{2}$ and $\pi^3(x) = \pi^0(x)$. The tensor
$\Theta^{\omega\varphi}$ is given in Ref.[11]: $\Theta^{\omega\varphi}
= g^{\omega\varphi} - (Z + 1/2)\gamma^{\omega}\gamma^{\varphi}$, where
the parameter $Z$ is arbitrary. The parameter $Z$ defines the ${\rm
\pi N \Delta}$ coupling off--mass shell of the $\Delta(1232)$
resonance. There is no consensus on the exact value of $Z$. From
theoretical point of view $Z=1/2$ is preferred [11].  Phenomenological
studies give only the bound $|Z| \le 1/2$ [13]. The empirical value of
the coupling constant $g_{\rm \pi N\Delta}$ relative to the coupling
constant $g_{\rm \pi NN}$ is $g_{\rm \pi N\Delta} = 2.12\,g_{\rm \pi
NN}$ [14]. As has been shown in Ref.[8] for the description of the
neutron--proton radiative capture for thermal neutrons the parameter
$Z$ should be equal to $Z = 0.438$. That is very close to $Z=1/2$.

For the subsequent calculations of the $\Delta\Delta$ component of the
deuteron it is useful to have the Lagrangian of the ${\rm \pi N
\Delta}$ interaction taken in the equivalent form
\begin{eqnarray}\label{label1.8}
{\cal L}_{\rm \pi N \Delta}(x) = \frac{g_{\rm \pi
N\Delta}}{2M_{\rm N}}\,\partial_{\varphi}\pi^a(x)\bar{N^c}(x)
\Theta^{\varphi\omega}\Delta^a_{\omega}(x)^c + {\rm h.c.},
\end{eqnarray}
where $\Delta^a_{\omega}(x)^c = C \bar{\Delta}^a_{\omega}(x)^T$. Now
we can proceed to the evaluation of the $\Delta\Delta$ component of
the deuteron.

\section{Effective ${\rm \Delta\Delta D}$ interaction}
\setcounter{equation}{0}

In the RFMD the existence of the $\Delta\Delta$ component of the
deuteron we can understand in terms of the coupling constant $g_{\rm
\Delta\Delta D}$ of the effective ${\rm \Delta\Delta D}$
interaction. 

In order to evaluate the Lagrangian of the effective ${\rm
\Delta\Delta D}$ interaction we have to obtain, first, the effective
Lagrangian of the transition N + N $\to$ $\Delta$ + $\Delta$. In the
RFMD this effective Lagrangian can be defined as follows [5,8]
\begin{eqnarray}\label{label2.1}
\int d^4z\,{\cal L}^{\rm NN \to \Delta\Delta}_{\rm eff}(z)&=&-\frac{g^2_{\rm \pi N\Delta}}{8M^2_{\rm N}}\int\!\!\!\int 
d^4x_1\,d^4x_2\,[\bar{\Delta}^a_{\alpha}(x_1)
\Theta^{\alpha\beta}N(x_1)]\,\nonumber\\
&&\times \frac{\partial}{\partial
x^{\beta}_1}\frac{\partial}{\partial
x^{\varphi}_1}[\delta^{ab}\Delta(x_1-x_2)]\,[\bar{N^c}(x_2)
\Theta^{\varphi\omega}\Delta^b_{\omega}(x_2)^c].
\end{eqnarray}
Since the transferred momenta are much less than the mass of the 
$\pi$--mesons, according to the prescription of the RFMD the Green
function of the $\pi$--mesons $\Delta(x_1-x_2)$ should be replaced by
the $\delta$--function [4--9]: $\Delta(x_1-x_2) =
\delta^{(4)}(x_1-x_2)/M^2_{\pi}$. This reduces  the effective
Lagrangian of the N + N $\to$ $\Delta$ + $\Delta$ transition to the form
\begin{eqnarray}\label{label2.2}
\int d^4z\,{\cal L}^{\rm NN \to \Delta\Delta}_{\rm eff}(z)&=&-\frac{g^2_{\rm \pi N\Delta}}{8M^2_{\pi}M^2_{\rm N}}\int\!\!\!\int 
d^4x_1\,d^4x_2\,[\bar{\Delta}^a_{\alpha}(x_1)
\Theta^{\alpha\beta}N(x_1)]\,\nonumber\\
&&\times \frac{\partial}{\partial
x^{\beta}_1}\frac{\partial}{\partial
x^{\varphi}_1}\delta^{(4)}(x_1-x_2)\,[\bar{N^c}(x_2)
\Theta^{\varphi\omega}\Delta^a_{\omega}(x_2)^c].
\end{eqnarray}
In terms of the Lagrangians of the ${\rm n p D}$ interaction and the N
+ N $\to$ $\Delta$ + $\Delta$ transition the Lagrangian of the
effective ${\rm \Delta\Delta D}$ interaction can be defined by [8]
\begin{eqnarray}\label{label2.3}
&&\int d^4x\,{\cal L}^{\rm \Delta\Delta D}_{\rm eff}(x) = -i\, \frac{g_{\rm
V}}{M^2_{\pi}}\,\frac{g^2_{\rm \pi N\Delta}}{4M^2_{\rm N}}\int
d^4x\,d^4x_1\,d^4x_2\,
D_{\mu}(x)\,\nonumber\\
&&[\bar{\Delta}^a_{\alpha}(x_1)
\Theta^{\alpha\beta}S_F(x-x_1)\gamma^{\mu}\tau_2S^c_F(x-x_2)
\Theta^{\varphi\omega}\Delta^a_{\omega}(x)^c]\,\frac{\partial}{\partial
x^{\beta}_1}\frac{\partial}{\partial
x^{\varphi}_1}\delta^{(4)}(x_1-x_2),
\end{eqnarray}
where $S_F(x-x_1)$ and $S^c_F(x-x_2)$ are the Green functions of the
free nucleon and anti--nucleon fields, respectively.

Such a definition of the contribution of the $\Delta\Delta$ component
to the deuteron is in agreement with that given by Niephaus {\it et
al.} [15] in the potential model appraoch (PMA).

In the large $N_C$ expansion [7] the Lagrangian of the effective ${\rm
\Delta\Delta D}$ interaction reduces itself to the local form and
reads
\begin{eqnarray}\label{label2.4}
{\cal L}^{\rm \Delta\Delta D}_{\rm eff}(x) =
\frac{1}{16\pi^2}\frac{g_{\rm V}}{M^2_{\pi}}\,\frac{g^2_{\rm \pi
N\Delta}}{4M^2_{\rm N}}
[\bar{\Delta}^a_{\alpha}(x)\,J^{\alpha\mu\omega}\tau_2
\Delta^a_{\omega}(x)^c]\,D_{\mu}(x),
\end{eqnarray}
where the structure function $J^{\alpha\mu\omega}$ is given by
the momentum integral
\begin{eqnarray}\label{label2.5}
J^{\alpha\mu\omega}=\int\frac{d^4k}{\pi^2i}\,
\Theta^{\alpha\beta}k_{\beta}\frac{1}{M_{\rm N} -
\hat{k}}\gamma^{\mu}\frac{1}{M_{\rm N} -
\hat{k}}k_{\varphi}\Theta^{\varphi\omega}.
\end{eqnarray}
The evaluation of the momentum integral in leading order in the large
$N_C$ expansion [7] gives
\begin{eqnarray}\label{label2.6}
J^{\alpha\mu\omega} = \frac{1}{6}\,\Lambda^3_{\rm D}\,M_{\rm N}\,\Theta^{\alpha\mu\omega}= \frac{1}{6}\,\Lambda^3_{\rm D}\,M_{\rm N}\,
\Theta^{\alpha\beta}\,(\gamma^{\mu}g_{\beta\varphi} -
g^{\mu}_{\beta}\gamma_{\varphi} -
g^{\mu}_{\varphi}\gamma_{\beta})\,\Theta^{\varphi\omega},
\end{eqnarray}
where $\Lambda_{\rm D} = 115.729\,{\rm Mev}$ [7] is a cut--off
restricting from above 3--momenta of the fluctuating nucleon fields
forming the physical deuteron [4,7].

Thus the Lagrangian of the effective ${\rm \Delta \Delta D}$
interaction reads
\begin{eqnarray}\label{label2.7}
&&{\cal L}^{\rm \Delta\Delta D}_{\rm eff}(x) = g_{\rm \Delta \Delta D}
[\bar{\Delta}^a_{\alpha}(x)\,\Theta^{\alpha\mu\omega}\tau_2
\Delta^a_{\omega}(x)^c]\,D_{\mu}(x)=\nonumber\\
&&=-\,i\,g_{\rm \Delta \Delta D}[\bar{\Delta}^-_{\alpha}(x)\Theta^{\alpha\mu\omega} \Delta^{++}_{\omega}(x)^c - \bar{\Delta}^{++}_{\alpha}(x)\Theta^{\alpha\mu\omega} \Delta^{-}_{\omega}(x)^c\nonumber\\
&& + \bar{\Delta}^+_{\alpha}(x)\Theta^{\alpha\mu\omega} \Delta^0_{\omega}(x)^c - \bar{\Delta}^0_{\alpha}(x)\Theta^{\alpha\mu\omega} \Delta^+_{\omega}(x)^c]\,D_{\mu}(x),
\end{eqnarray}
where the effective coupling constant $g_{\rm \Delta \Delta D}$ is
defined by
\begin{eqnarray}\label{label2.8}
g_{\rm \Delta \Delta D} = g_{\rm V}\,\frac{g^2_{\rm \pi
NN}}{384\pi^2}\,\frac{g^2_{\rm \pi
N\Delta}}{g^2_{\rm \pi
NN}}\frac{\Lambda^3_{\rm D}}{M^2_{\pi}M_{\rm N}}.
\end{eqnarray}
On--mass shell of the $\Delta(1232)$ resonance, i.e. in the case of
the PMA [1,15], the contribution of the parameter $Z$ vanishes and the
effective ${\rm \Delta \Delta D}$ interaction acquires the form
\begin{eqnarray}\label{label2.9}
&&{\cal L}^{\rm \Delta\Delta D}_{\rm eff}(x) = 
g_{\rm \Delta \Delta
D}\,g^{\alpha\beta} [\bar{\Delta}^a_{\alpha}(x)\,\gamma^{\mu}\tau_2
\Delta^a_{\beta}(x)^c]\,D_{\mu}(x)=\nonumber\\ 
&&=-\,i\,g_{\rm \Delta
\Delta D}\,g^{\alpha\beta}\,[\bar{\Delta}^-_{\alpha}(x)\gamma^{\mu}
\Delta^{++}_{\beta}(x)^c -
\bar{\Delta}^{++}_{\alpha}(x)\gamma^{\mu}
\Delta^{-}_{\beta}(x)^c\nonumber\\ && +
\bar{\Delta}^+_{\alpha}(x)\gamma^{\mu}
\Delta^0_{\beta}(x)^c -
\bar{\Delta}^0_{\alpha}(x)\gamma^{\mu}
\Delta^+_{\beta}(x)^c]\,D_{\mu}(x).
\end{eqnarray}
The total probability of the existence of the $\Delta\Delta$ component
in the deuteron we can estimate by the quantity
\begin{eqnarray}\label{label2.10}
P(\Delta\Delta) = 2\,\frac{g^2_{\rm \Delta \Delta D}}{g^2_{\rm V}} = 0.08\,\%.
\end{eqnarray}
Our theoretical prediction agrees good with the recent experimental
estimate of the upper limit $P(\Delta\Delta) < 0.4\%$ at 90$\%$ of
CL [16] quoted by Dymarz and Khanna [1].

\section{Conclusion}

The theoretical estimate of the $\Delta\Delta$ component of the
deuteron obtained in the RFMD agrees good with the experimental upper
limit. Indeed, for the $\Delta(1232)$ resonance on--mass shell [1,15] we
predict $P(\Delta\Delta) = 0.08\,\%$ whereas experimentally
$P(\Delta\Delta)$ is restricted by $P(\Delta\Delta) < 0.4\%$ at 90$\%$ of
CL [16]. 

Off--mass shell of the $\Delta(1232)$ resonance, where the parameter
$Z$ should contribute, our prediction for $P(\Delta\Delta)$ can be
changed, of course. Moreover, the contribution of the $\Delta\Delta$
component to different processes and quantities can be
different. However, we would like to emphasize that in the RFMD by
using the effective ${\rm \Delta \Delta D}$ interaction determined by
Eq.(\ref{label2.7}) one can calculate the contribution of the
$\Delta\Delta$ component of the deuteron to any low--energy process
with the deuteron in the initial or final state. These calculations are
rather complicated and go beyond the scope of this letter.

In our approach we do not distinguish contributions of the
$\Delta\Delta$--pair with a definite orbital momentum ${^3}{\rm
S^{\Delta\Delta}_1}$, ${^3}{\rm D^{\Delta\Delta}_1}$ and so on to the
effective ${\rm \Delta \Delta D}$ interaction Eq.(\ref{label2.7}).
The obtained value of the probability $P(\Delta\Delta)$ should be
considered as a sum of all possible states with a certain orbital
momentum.

Our prediction $P(\Delta\Delta) = 0.08\,\%$ does not contradict to the
predictions obtained in the PMA by Dymarz and Khanna [1]. In fact, as
has been stated by Dymarz and Khanna [1] their results agree with the
experimental estimate of the upper limit of the total probability
$P(\Delta\Delta)$ given by Allia {\it et al.} [16]. Unlike our
approach Dymarz and Khanna have given a percentage of the
probabilities of different states ${^3}{\rm S^{\Delta\Delta}_1}$,
${^3}{\rm D^{\Delta\Delta}_1}$ and so to the wave function of the
deuteron. In our approach the deuteron couples to itself and other
particles through the one--baryon loop exchanges. The Lagrangian of
the effective ${\rm \Delta\Delta D}$ interaction given by
Eq.(\ref{label2.7}) defines completely the contribution of the
$\Delta\Delta$ intermediate states to baryon--loop exchanges. The
decomposition of the effective ${\rm \Delta\Delta D}$ interaction
according to the $\Delta\Delta$ states with a certain orbital momentum
should violate Lorentz invariance for the evaluation of the
contribution of every state. In the RFMD this can lead to incorrect
results [7--9]. The relativistically covariant procedure of the
decomposition of the interactions like the ${\rm \Delta\Delta D}$ one
in terms of the states with a certain orbital momenta is now in
progress in the RFMD. However, a smallness of the contribution of the
$\Delta\Delta$ component in the deuteron obtained in the RFMD makes
such a decomposition applied to the ${\rm \Delta\Delta D}$ interaction
meaningless to some extent due to impossibility to measure the terms 
separately.

\section{Acknowledgement}

We are grateful to Prof. W. Plessas for discussions stimulated this
investigation.


\newpage

\begin{thebibliography}{9}
\bibitem{[1]}
R. Dymarz and F. C. Khanna,
Nucl. Phys. A516 (1990) 549.
\bibitem{[2]}
{\it QUARKS AND NUCLEI}, ed W. Weise, World  Scientific, Singapore, 1989.
\bibitem{[3]} {\it MESONS IN NUCLEI}, ed. M. Rho and D. H. Wilkinson,
Noth--Holland, Amsterdam, 1979.
\bibitem{[4]}
A. N. Ivanov, N. I. Troitskaya, M. Faber and H. Oberhummer,
Phys. Lett. B361 (1995) 74; Nucl. Phys. A617 (1997) 414,
{\it ibid.} A625 (1997) 896 (Erratum).
\bibitem{[5]}
A. N. Ivanov, N. I. Troitskaya, H. Oberhummer and M. Faber,
Z. Phys. A358 (1997) 81.
\bibitem{[6]} 
A. N. Ivanov, H. Oberhummer, N. I. Troitskaya and M. Faber, 
{\it Solar proton burning, photon and anti--neutrino
disintegration of the deuteron in the relativistic field theory model
of the deuteron}, nucl--th/9810065, October 1998; 
{\it Solar neutrino processes in the relativistic field theory model 
of the deuteron}, nucl--th/9811012, November 1998.
\bibitem{[7]}
 A. N. Ivanov, H. Oberhummer, N. I. Troitskaya and M. Faber, 
{\it The relativistic field theory model of the deuteron
from low--energy QCD}, nucl--th/9908029, August 1999.
\bibitem{[8]} 
A. N. Ivanov, H. Oberhummer, N.I. Troitskaya $\&$ M. Faber, 
{\it Neutron--proton radiative capture, photo--magnetic and
anti--neutrino disintegration of the deuteron in the relativistic
field theory model of the deuteron}, nucl--th/9908080, August 1999.
\bibitem{[9]} 
A. N. Ivanov, H. Oberhummer, N.I. Troitskaya $\&$
M. Faber, {\it Solar proton burning, neutrino disintegration of the
deuteron and the pep process in the relativistic field theory model of
the deuteron}, nucl--th/9908080, August 1999.
\bibitem{[10]}
W. Rarita and J. Schwinger,
Phys. Rev. 60 (1941) 61.
\bibitem{[11]}
L. M. Nath, B. Etemadi and J. D. Kimel,
Phys. Rev. D3 (1971) 2153.
\bibitem{[12]} 
J. Kambor, 
{\it The $\Delta(1232)$ as an Effective
Degree of Freedom in Chiral Perturbation Theory}, Talk given at the
Workshop on Chiral Dynamics, 1997 Mainz, Germany, September 1--5,
1997; hep--ph/9711484 26 November 1997.
\bibitem{[13]} K. Kabir, T. K. Dutta, Muslema Pervin and L. M. Nath,
{\it The Role of $\Delta(1232)$ in Two--pion Exchange Three--nucleon
Potential}, hep--th/9910043, October 1999.
\bibitem{[14]}
V. Bernard, N. Kaiser and Ulf--G. Meissner,
Int. J. Mod. Phys. E4 (1995) 193 and references therein.
\bibitem{[15]}
G. H. Niephaus, M. Gari and B. Sommer,
Phys. Rev. C20 (1979) 1096.
\bibitem{[16]} 
D. Allasia {\it et al.}, Phys. Lett. B174 (1986) 450.
\end{thebibliography}
\end{document}